\documentclass{aa}
\usepackage{graphics}
\usepackage{natbib}
\newcommand{\Msun}{M$_{\sun}$\ }
\newcommand{\Lsun}{L$_{\sun}$\ }

\newcommand{\Lstar}{L$_{\star}$\ }

\newcommand{\CS}{Cloud Size\ }

\newcommand{\aas}{A{\&}AS\ }
\newcommand{\annrev}{ARA{\&}A\ }
\begin{document}
\title{Classical T Tauri stars as sources of parsec\,-\,scale optical outflows}
\author{F. McGroarty \inst{1}
\and T.P. Ray \inst{1}}
\offprints{F. McGroarty, fmcg@cp.dias.ie}
\institute{Dublin Institute for Advanced Studies, 5 Merrion Square, Dublin 2, Ireland}
\date{Received date ;accepted date}

\abstract{
Previous studies of young stellar objects (YSOs) have uncovered a number of associated parsec\,-\,scale
optical outflows, the majority of which are driven by low\,-\,mass, embedded Class I sources. Here we 
examine more evolved Classical T Tauri stars (CTTSs), i.e. Class II sources, to determine whether these
are also capable of driving parsec\,-\,scale outflows. Five such sources are presented here - CW Tau, 
DG Tau, DO Tau, HV Tau C and RW Aur, all of which show optical evidence for outflows of the order of 
1pc (24$'$ at the distance of Taurus\,-\,Auriga). These sources were previously known only to drive 
``micro\,-\,jets'' or small\,-\,scale outflows $\la$ 1\arcmin\ in length. A parsec\,-\,scale outflow 
from a less evolved source (DG Tau B) which was noted in the course of this work is also included here.
Examination of the five newly discovered large\,-\,scale outflows from CTTSs shows that they have 
comparable morphologies, apparent dynamical timescales and degrees of collimation to those from less 
evolved sources. There is also strong evidence that these outflows have blown out of their parent 
molecular clouds. Finally we note that the ``fossil record'' provided by these outflows suggests their 
sources could have undergone FU Orionis\,-\,type outbursts in the past.
 
}

\maketitle
~~~~~~~~~~~~~~~~~~~~~~~~~~~~~~~~~~~~~~~~~~~~~~~~~~~~~~~~~~~~~~~~~~~~~~~~~~~~~~~~~~~~~~~~~~~~~~~~~~~~~~~
\section{Introduction}
Herbig\,-\,Haro (HH) objects are the optically visible tracers of mass outflow from YSOs and are
therefore ultimately powered by accretion \citep{Cabrit90,Hartigan95}. Over the years many bipolar HH 
outflows have been observed and most were found to be driven by embedded, low\,-\,mass sources of $\la$
1\Msun$\!\!$. Initially it was assumed that their lengths were only a fraction of a parsec, however the
discovery of a $\sim$ 1.4pc long outflow from RNO\,43 \citep{Ray87} hinted that this may not always be 
the case. In the mid 1990's it was realised that many of these outflows can have projected lengths much
greater than 1pc \citep{Bally96,Eisloffel97,Reipurth97}, reaching up to $\sim$ 11pc. Some well known 
examples are the HH\,34 outflow consisting of HH\,33, HH\,40, HH\,85, HH\,126, HH\,34N, HH\,34, HH\,34X,
HH\,173, HH\,86, HH\,87 and HH\,88 which is 3pc in projected length \citep{Bally94}; the 5.9pc long 
HH\,401, HH\,1, HH\,2 and HH\,402 outflow \citep{Ogura95}; and the HH\,113\,/\,HH\,111\,/\,HH\,311 
outflow at 7.7pc in length \citep{Reipurth97}.  

It is not surprising that such outflows can attain these lengths when we consider that they have 
tangential velocities of between 50 -- 200 kms$^{-1}$ \citep{Devine97,Reipurth97} and the outflow 
phase, for even (low\,-\,mass) Class I sources, lasts at least 10$^5$ years. In reality, it should be 
{\em expected} that most will attain parsec\,-\,scale lengths. The main observational hindrance in the 
past to observing them was the relatively small fields of view offered by most CCD cameras. With the 
advent of large CCD mosaics more and more parsec\,-\,scale outflows have been discovered.

The morphology of parsec\,-\,scale outflows yields valuable information about their driving 
sources. They are, in effect, a fossil record of the mass\,-\,loss history of their source over their 
dynamical timescales ($\sim$10$^3$ to $\sim$10$^5$ yr). They suggest, for example, quiescent 
phases between periods of violent mass ejection that give rise to the large HH complexes we see today.  
The morphology of an outflow can also indicate whether it is precessing and, if so, the rate of 
precession. 

As mentioned earlier many of the parsec\,-\,scale outflows that have been observed to date are driven 
by young, Class I, low\,-\,mass YSOs. The classification scheme used here is based on the shape of
the spectral energy distribution (SED) of the YSO from 10$\mu$m to 100$\mu$m \citep{Lada84,Lada87}. The 
SED of a YSO can be modelled as an approximate blackbody with an infrared excess longwards of 2$\mu$m 
due to circumstellar dust and gas. The infrared excess is very strong in the young, embedded Class I 
sources and is almost non\,-\,existent in the most evolved Class III sources. Here we observed a number
of Class II low\,-\,mass sources -- CTTSs. These CTTSs were not previously associated with 
parsec\,-\,scale outflows; in fact many were only known to drive ``micro\,-\,jets'' of the order of 
$\sim$ 5\arcsec\ to 40\arcsec\ ($\la$ 0.03pc at a distance of 140pc to the Taurus\,-\,Auriga Cloud). 
Although outflows from these more evolved sources are not nearly as spectacular as those from Class I 
YSOs, their sources are no longer surrounded by significant amounts of dust and so their outflows can
be traced right back to the origin. These Class II sources are still actively accreting and ejecting 
matter, albeit at rates 10\,-\,100 times smaller than Class I sources \citep{Hartigan95}. In this 
paper, we present a number of parsec\,-\,scale outflows from CTTSs and we investigate whether these 
CTTSs show evidence for having undergone FU Orionis-like outbursts, based on the fossil record of their
outflows.

Details about the observations are given in Section 2. In Section~\ref{sec-resultsI} we report the 
discovery of parsec\,-\,scale outflows from five CTTSs in the Taurus\,-\,Auriga Cloud, at a distance of 
140pc \citep{Elias78,Wichmann98}. We also include serendipitous observations of a parsec\,-\,scale 
outflow from a less evolved Class I source. These results are discussed in Section~\ref{sec-discussion} 
and our conclusions are presented in Section~\ref{sec-conclusions}. 

\section{Observations}

Our data was acquired using the Wide Field Camera (WFC) on the 2.5m Isaac Newton Telescope at El 
Observatorio del Roque de los Muchachos (La Palma, Canary Islands). The WFC consists of four 
thin\,-\,coated EEV CCDs each with 2048 $\times$ 4100, 15$\mu$m $\times$ 15$\mu$m pixels. One pixel 
projects to 0\farcs33 on the sky. Three of the CCDs are positioned from north to south with their long 
axes adjoining. The fourth is attached to the west to form a square (34\farcm2 wide) with its 
northwestern corner missing.

Our images were taken during two separate observing runs, the first between the 10$^{\rm th}$ and 
13$^{\rm th}$ of February 2001, and the second between the 24$^{\rm th}$ and 27$^{\rm th}$ of November 
2003. HH objects were identified using a number of narrowband emission line filters: 
H$\alpha$($\lambda_c$ = 6568\AA, $\Delta\lambda$(FWHM) = 95\AA ) and  [SII]($\lambda_c$ = 6725\AA, 
$\Delta\lambda$(FWHM) = 80\AA ). To distinguish HH emission from reflection nebulosity, we also took 
broadband images in I and V. Seeing in the images taken in February 2001 was moderate at 
1\arcsec\,--\,2\arcsec\ as measured from the images. The seeing in the November 2003 images was better 
at 0\farcs9\,--\,1\farcs1. Exposure times for the narrowband and broadband images were typically 30 and 
10 minutes respectively. The data were reduced using standard IRAF reduction procedures.
The sources observed here are CTTSs with previously known ``micro\,-\,jets'' or short outflows of 
$\la$ 1\arcmin. All are in the Taurus\,-\,Auriga cloud, chosen for its abundance of CTTSs. Seven such 
sources were observed, with 5 of them -- CW Tau, DG Tau, DO Tau, HV Tau C and RW Aur -- revealing 
extended outflows of the order of 1pc. No {\it extended} optical emission was found in the UY Aur and 
DP Tau outflows.

\begin{table}
\begin{tabular}{llll}
\hline \hline
Object       &Source            &$\alpha$(J2000)      &$\delta$(J2000)         \\ \hline
HH\,826\,A   &CW Tau            &04$^h$14$^m$17.8$^s$ &+28\degr 10$'$40$''$    \\ 
HH\,826\,B   &CW Tau            &04$^h$14$^m$19.8$^s$ &+28\degr 09$'$52$''$    \\ 
HH\,827      &CW Tau            &04$^h$14$^m$15.1$^s$ &+28\degr 03$'$55$''$    \\ 
HH\,826\,C   &CW Tau            &04$^h$14$^m$15.3$^s$ &+28\degr 11$'$39$''$    \\ 
HH\,828      &CW Tau            &04$^h$14$^m$10.3$^s$ &+28\degr 14$'$54$''$    \\ 
HH\,829\,A   &CW Tau            &04$^h$14$^m$03.0$^s$ &+28\degr 25$'$36$''$    \\ 
HH\,829\,B   &CW Tau            &04$^h$14$^m$04.8$^s$ &+28\degr 26$'$53$''$    \\ 
HH\,829\,C   &CW Tau            &04$^h$14$^m$07.3$^s$ &+28\degr 27$'$37$''$    \\ 
HH\,830\,A   &DG Tau            &04$^h$27$^m$37.3$^s$ &+26\degr 12$'$27$''$    \\ 
HH\,830\,B   &DG Tau            &04$^h$27$^m$44.8$^s$ &+26\degr 12$'$56$''$    \\ 
HH\,830\,C   &DG Tau            &04$^h$27$^m$51.7$^s$ &+26\degr 15$'$33$''$    \\ 
HH\,831\,A   &DO Tau            &04$^h$39$^m$13.2$^s$ &+26\degr 13$'$48$''$    \\ 
HH\,831\,B   &DO Tau            &04$^h$39$^m$15.2$^s$ &+26\degr 13$'$55$''$    \\ 
HH\,832      &DO Tau            &04$^h$39$^m$02.0$^s$ &+26\degr 12$'$21$''$    \\ 
HH\,833      &HV Tau C          &04$^h$38$^m$44.0$^s$ &+26\degr 14$'$42$''$    \\ 
HH\,834      &\ \ \ \ \ \   ?   &04$^h$39$^m$05.9$^s$ &+26\degr 03$'$23$''$    \\ 
HH\,835      &RW Aur            &05$^h$07$^m$30.4$^s$ &+30\degr 27$'$11$''$    \\ 
HH\,836\,A   &DG Tau B          &04$^h$27$^m$13.5$^s$ &+26\degr 04$'$16$''$    \\ 
HH\,836\,B   &DG Tau B          &04$^h$27$^m$19.7$^s$ &+26\degr 03$'$07$''$    \\ 
HH\,837      &DG Tau B          &04$^h$27$^m$44.8$^s$ &+26\degr 00$'$49$''$    \\ 
HH\,838      &\ \ \ \ \ \   ?   &04$^h$26$^m$56.4$^s$ &+26\degr 05$'$58$''$    \\ 
HH\,839      &\ \ \ \ \ \   ?   &04$^h$27$^m$43.8$^s$ &+26\degr 04$'$35$''$    \\ \hline
\end{tabular}
\caption{Positions of the new HH objects found in this survey and their {\em probable} sources.}
\label{HHPositions}
\end{table}

\section{Results}
\label{sec-resultsI}
\subsection{CW Tau}
\label{sec-cwtau}
CW Tau has a spectral type of K3  \citep{Cohen79}, with M$_{\star}$ = 1.40 M$_{\odot}$ and L$_{\star}$ 
= 2.6 L$_{\odot}$ \citep{GomezdeCastro93}. A ``micro\,-\,jet'' (HH\,220) propagating southeast from 
this source was discovered in optical images by \cite{GomezdeCastro93}, with a gap of $\sim$ 1\farcs3 
between the source and the jet. A full opening angle of 3.3\degr\ is derived for this blueshifted jet 
at 3\arcsec\ from the source \citep{Dougados00}. Long\,-\,slit spectroscopic observations by 
\cite{Hirth94b} showed the outflow to extend at least 4\arcsec\ - 6\arcsec\ on either side of the 
source. The blueshifted HH\,220 outflow is at a position angle (P.A.) of 144\degr\ $\pm$ 2\degr\ with 
respect to CW Tau \citep{GomezdeCastro93,Dougados00} and we estimate the redshifted jet to be at $\sim$ 
329\degr\ with respect to CW Tau from the [SII] images of \cite{Dougados00}. 

\begin{figure}
\resizebox{\hsize}{!}{\includegraphics{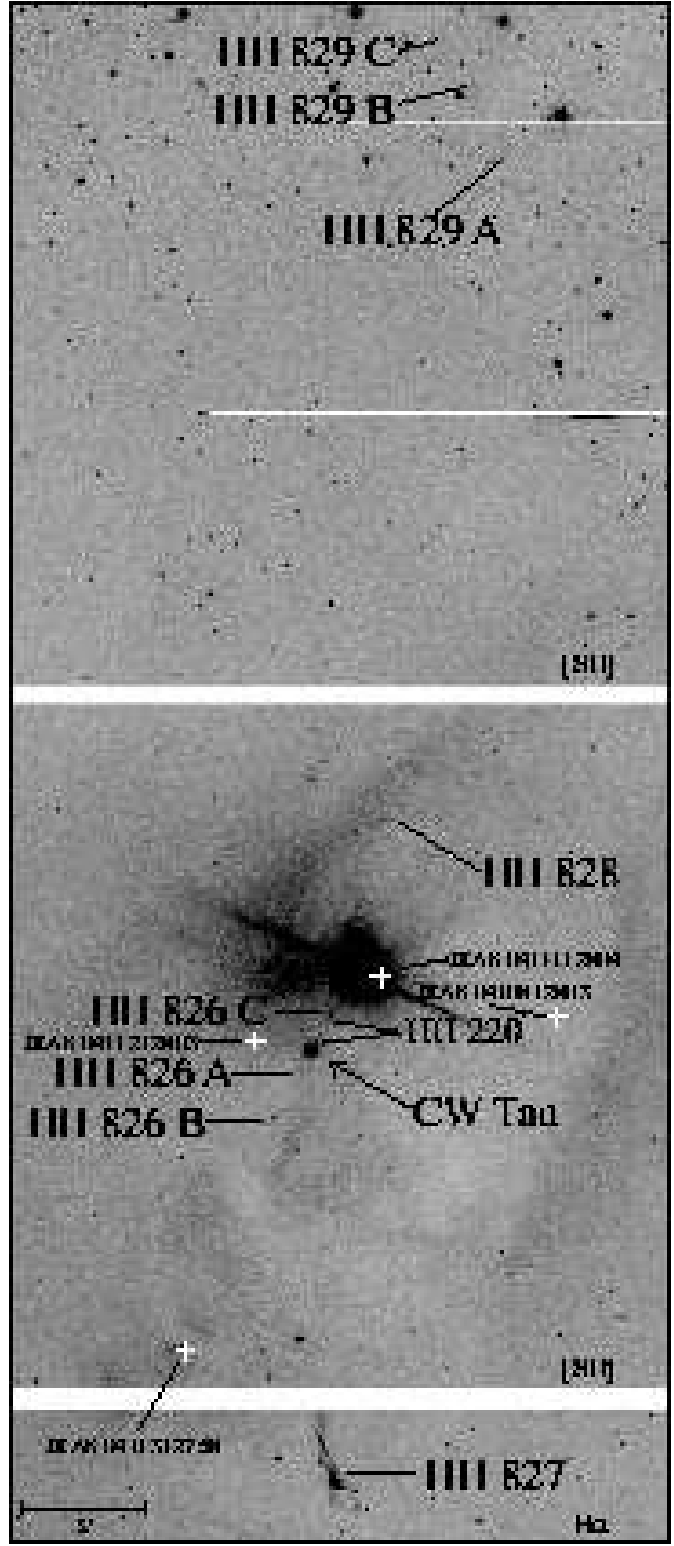}}
\caption{{\it CW Tau :\/} Mosaic of the entire CW Tau outflow. The top and middle frames
are in [SII] and the bottom frame is in H$\alpha$ as HH\,827 is much stronger in H$\alpha$. All known 
IRAS sources in the region are marked with white crosses. The individual knots can be seen more clearly
in Figs.\ \ref{cwtau_center_hh220} to\ \ref{cwtau_hh829}. Any information in the gaps between the CCDs 
(white strips) is lost. The northern edge of this image is the northern edge of the field of view of 
the CCD Mosaic. Note that the linear object to the east of HH\,829\,B is an asteroid trail. For all 
images in this paper, North is up and West is to the right.}
\label{cwtau_flow}
\end{figure}

\begin{figure}
\resizebox{\hsize}{!}{\includegraphics{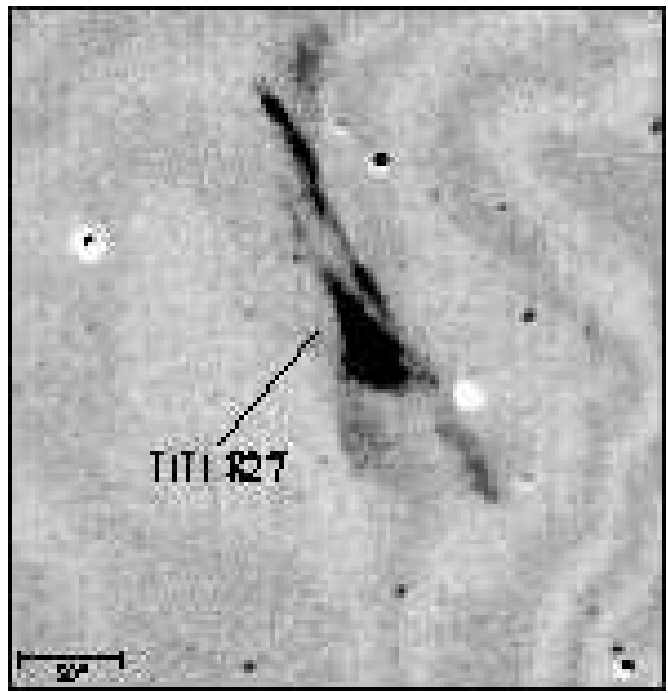}}
\caption{{\it HH\,827 H$\alpha$ :\/} Continuum subtracted H$\alpha$ image of HH\,827 to
the south of CW Tau. The trail of emission to the northeast is clearly seen and the fainter trail
to the southwest extends for $\sim$ 31\arcsec\ from the southern end of the bright knot.}
\label{cwtau_hh827_ctm}
\end{figure}

Our wide field images of the region around CW Tau (Fig.\ \ref{cwtau_flow}) reveal that this outflow is 
much more extended than just the HH\,220 bipolar jet -- see Table\ \ref{cwtau_objects} for details of 
the newly discovered objects in the CW Tau outflow. Two knots are found to the south of CW Tau 
at 22\arcsec\ (HH\,826\,A) and 1\farcm27 (HH\,826\,B). HH\,826\,A is only seen in [SII] emission while 
HH\,827\,B is seen in both [SII] and H$\alpha$. Both are at a P.A. of $\sim$ 153\degr\ with respect to 
the source and are reasonably well aligned with the blueshifted HH\,220 jet. Further out at 6\farcm1 is
HH\,827 at a P.A. of 184\degr\ with respect to CW Tau. Precession of the outflow may explain this 
directional change. HH\,827 consists of a bright knot with a trail of emission stretching to the 
northeast and a fainter trail of emission to the southwest (see Fig.\ \ref{cwtau_hh827_ctm}) and is 
much brighter and more extended in H$\alpha$ than in [SII]. The total length of this object in 
H$\alpha$ is $\sim$ 1\farcm8. It is possible that HH\,827 is not driven by CW Tau in which case 
IRAS 04113+2758 and IRAS 04112+2803 to the northeast and north of HH\,827 (marked on 
Fig.\ \ref{cwtau_flow}) are candidate sources, however the trail of emission from HH\,827 doesn't point
back to either of these. If HH\,827 is driven by CW Tau then the projected length of the blueshifted 
outflow is 7\farcm08 (0.29pc). 

\begin{table}
\begin{tabular}{llccl}
\hline \hline
Object      &Source          &Angular        &P.A.$^b$        &Spatial        \\ 
            &               &Separation$^a$  &/\degr          &Extent$^c$     \\ \hline
HH\,826\,A  &CW Tau         &0\farcm37       &153               &2\arcsec $\times$ 2\arcsec \\
HH\,826\,B  &CW Tau         &1\farcm27       &153               &2\arcsec $\times$ 5\arcsec \\
HH\,827     &CW Tau         &6\farcm2        &184               &1\farcm8 $\times$ 16\arcsec \\
HH\,826\,C  &CW Tau         &0\farcm77       &326               &4\arcsec $\times$ 3\arcsec \\
HH\,828     &CW Tau         &4\farcm52$^c$  &342$^c$            &3\arcsec $\times$ 2\arcsec (e)$^d$\\
            &CW Tau         &4\farcm27$^c$  &338$^c$            &2\arcsec $\times$ 3\arcsec (m)$^d$\\
            &CW Tau         &4\farcm33$^c$  &334$^c$            &5\arcsec $\times$ 3\arcsec (w)$^d$\\
HH\,829\,A  &CW Tau         &14\farcm9       &348               &29\arcsec $\times$ 5\arcsec \\
HH\,829\,B  &CW Tau         &16\farcm12      &351               &3\arcsec $\times$ 4\arcsec \\
HH\,829\,C  &CW Tau         &16\farcm8       &353               &3\arcsec $\times$ 6\arcsec \\ \hline
\end{tabular}
\caption{Angular separations, P.A.s and spatial extent of newly discovered HH objects in the CW Tau 
region.
\newline $^a$ : Angular distance from the presumed source.
\newline $^b$ : P.A. with respect to the presumed source.
\newline $^c$ : Width and length of the object respectively.                          
\newline $^d$ : The three knots in HH\,828 (see Fig.\ \ref{cwtau_center_hh220}) are labelled here as 
the eastern\,-\,most knot (e), the middle knot (m) and the western\,-\,most knot (w).}
\label{cwtau_objects}
\end{table}

\begin{figure}
\resizebox{\hsize}{!}{\includegraphics{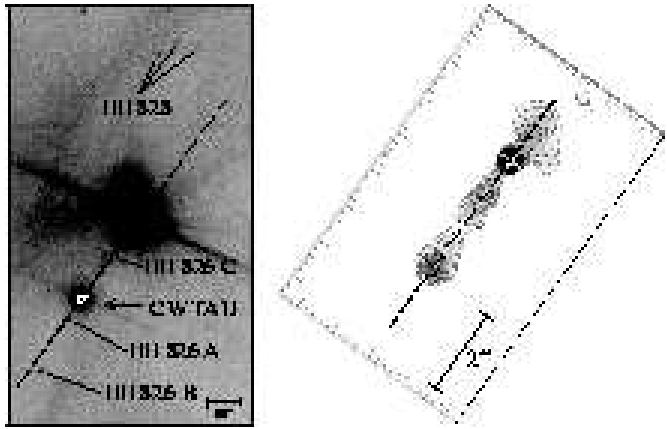}}
\caption{{\it CW Tau :\/} {\it Left :\/} Our [SII] image showing CW Tau and the newly 
discovered HH\,826 and HH\,828 knots. The small white box superimposed on CW Tau marks the region shown
on the right. {\it Right :\/} CW Tau and the HH220 jet taken from \cite{Dougados00}. A dotted line 
shows the P.A. of 144\degr\ determined by \cite{Dougados00} and a similar line at 144\degr\ is 
superimposed on our [SII] image to show the slight change in direction of the outflow.} 
\label{cwtau_center_hh220}
\end{figure}

\begin{figure}
\resizebox{\hsize}{!}{\includegraphics{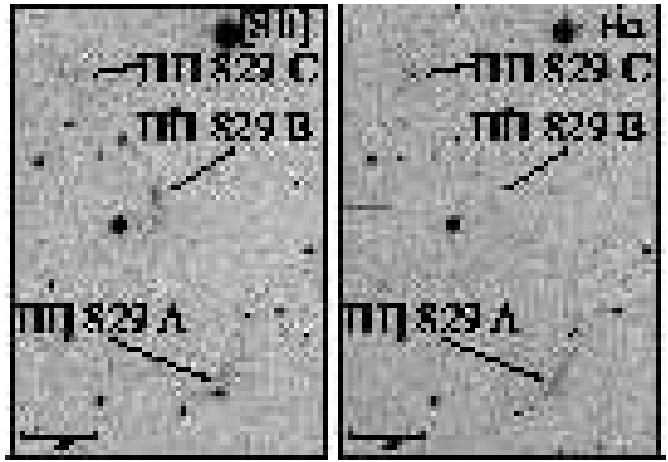}}
\caption{{\it HH\,829 :\/} Image of HH\,829\,A, B and C to the north of CW Tau in [SII] (left) and 
H$\alpha$ (right). HH\,829\,C is stronger in H$\alpha$ emission and HH\,829\,B is 
stronger in [SII] emission, while HH\,829\,A is comparable in both.}
\label{cwtau_hh829}
\end{figure}

Our [SII] images show that the redshifted HH\,220 jet extends to $\sim$ 46\arcsec\ from CW Tau at a 
P.A. of 326\degr\ (Fig.\ \ref{cwtau_flow}). The jet first extends to $\sim$ 9\arcsec\ from the 
source, then there is a gap of almost 20\arcsec\ where the jet is too faint to be seen. It becomes 
visible again for a distance of $\sim$ 17\arcsec\ before terminating in the bright knot HH\,826\,C. 
Further out a trio of knots (HH\,828) are found at $\sim$ 4\farcm3 from CW Tau and are only seen in 
[SII] (Figs.\ \ref{cwtau_flow} \& \ref{cwtau_center_hh220}). The most western knot is at a P.A. of 
334\degr\ from CW Tau and the most eastern is at 342\degr. There are three known IRAS sources in the 
vicinity of HH\,828 (marked in Fig.\ \ref{cwtau_flow}) and it is possible that the source of the 
HH\,828 knots is one of these or CW Tau itself. Proper motion studies would help distinguish between 
these possibilities.

There is a distance of $\sim$ 10\farcm8 (0.44pc) between HH\,828 and the next HH object along this 
direction, HH\,829\,A (Figs.\ \ref{cwtau_flow} \& \ref{cwtau_hh829}). HH\,829\,A, B and C are 
14\farcm9, 16\farcm12 and 16\farcm8 at 348\degr, 351\degr\ and 353\degr\ from CW Tau respectively. The 
edge of this complex is approximately 37\arcsec\ from the northern edge of our field of view so it is 
possible that further emission is present beyond this. While HH\,829\,A is comparable in brightness in 
both H$\alpha$ and [SII] emission, knots B and C are brighter in [SII] and H$\alpha$ respectively 
(Fig.\ \ref{cwtau_hh829}). IRAS 04111+2820 (slightly outside the field of view of 
Fig.\ \ref{cwtau_flow}) is $\sim$ 1\farcm5 at a P.A. of 78\degr\ from HH\,829\,C but is highly unlikely
to be driving this set of co-linear knots given their position with respect to this source. The 
projected length of this assumed CW Tau redshifted outflow is 0.69pc (16\farcm8). This gives a total 
projected length of the CW Tau blue\,- and redshifted outflows of $\sim$ 0.98pc. 

The variation in P.A. in the blueshifted outflow is 40\degr\ and is 24\degr\ in the redshifted outflow.
This change in direction gives the extended outflow an inverted `S' shape. Similar morphologies are 
seen in many large\,-\,scale outflows from less evolved low\,-\,mass sources (see 
Section~\ref{sec-morphology}) and suggests outflow precession. The change in direction is approximately
symmetrical about CW Tau which substantiates our suggestion that both HH\,827 and HH\,829 may be part 
of this outflow. CW Tau is surrounded by a dark cloud (Fig.\ \ref{cwtau_flow}) so the majority of the 
outflow is presumably obscured by this cloud. It is interesting to note however that the more distant 
objects in this outflow, HH\,827 and HH\,829, are found at the cloud edges.

\subsection{The DG Tau region}
\label{sec-dgtau}

\begin{figure*}[t]
\resizebox{\hsize}{!}{\includegraphics{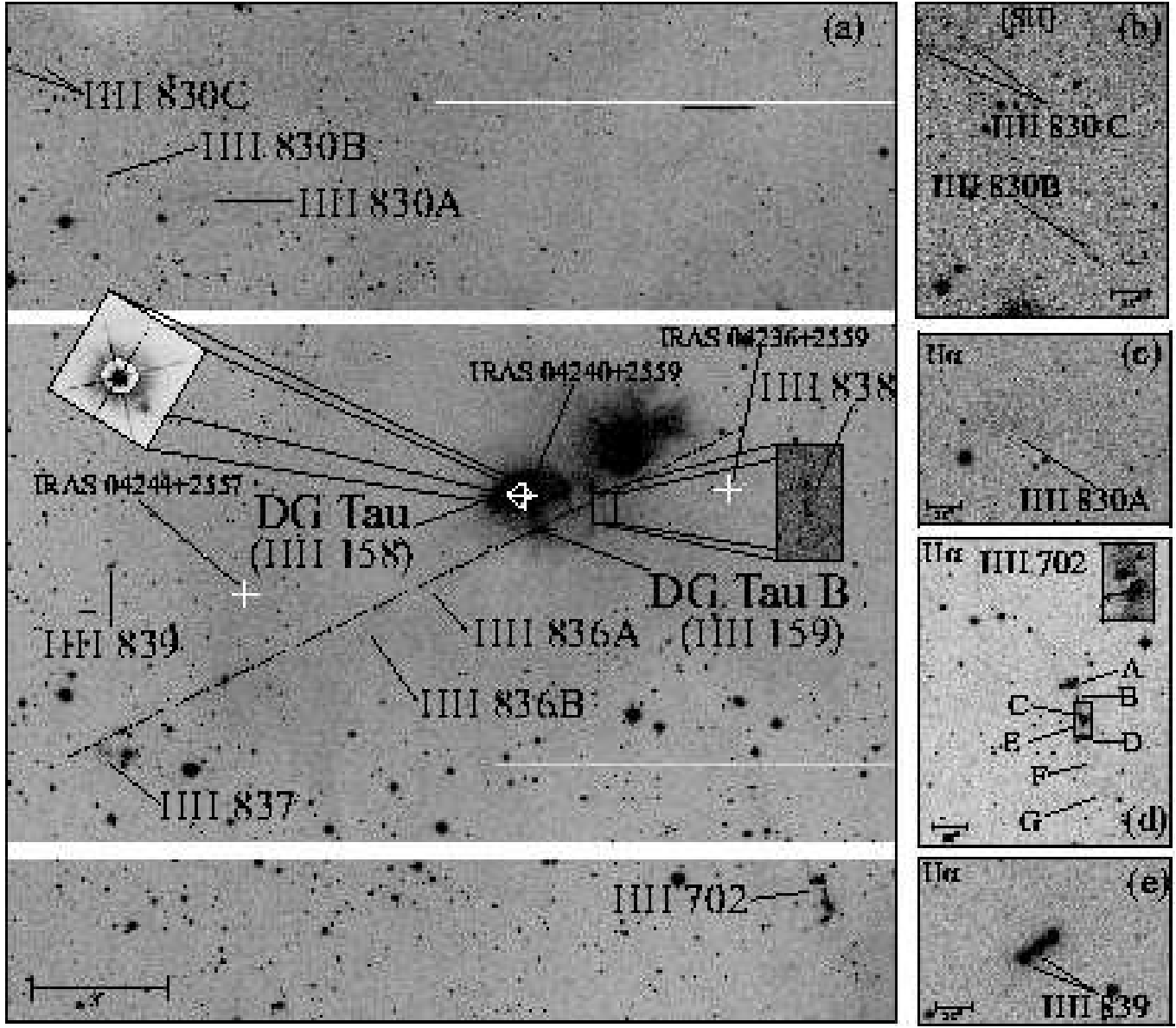}}
\caption{{\it DG Tau :\/} (a) Mosaic image of the region surrounding DG Tau and DG Tau B in H$\alpha$ :
The new objects discovered in both the DG Tau outflow (HH\,830) and the DG Tau B outflow (HH\,836 and 
HH\,837) are indicated here. HH\,838 and HH\,839 appear to be separate outflows in this region. A HST 
image of HH\,158 is included here as an inset and has been rotated such that north is up. This inset is
21\farcs7\ by 22\farcs05. The dotted line through the centre of the image is at 116\degr\ and marks the
blueshifted DG Tau B outflow. The three known IRAS sources in the region are marked with white crosses.
(b) HH\,830\,B and C in [SII] : HH\,830\,C can be seen here to be two distinct regions of emission, and
HH\,830\,B is a small knot. Note that the left hand edge of the image is actually the eastern edge of 
our field of view of the CCD mosaic so it is quite possible that HH\,830\,C is more extended. (c) 
HH\,830\,A in H$\alpha$ : HH\,830\,A is a faint, linear emission object extending $\sim$ 28\arcsec\ and 
dips southwards towards the western edge for another 16\arcsec. (d) HH\,702 in H$\alpha$ : An image 
taken in February 2001 is used here as it shows the whole of HH\,702 in one CCD field. Knots A, B, C 
and D (discovered by \cite{Sun03}) can clearly be seen. The newly discovered Knot E and fainter 
Knots F and G are also marked here. There is a star to the north of Knot E but this knot can be seen 
clearly in the H$\alpha$ continuum subtracted image (inset). (e) HH\,839 in H$\alpha$.} 
\label{dgtau_all}
\end{figure*}

DG Tau is a low luminosity star with \Lstar $\simeq$ 8\Lsun \citep{Cohen79} and was one of the first
TTSs to be associated with an optical jet (HH\,158), noted by \cite{Mundt83}. HST\,/\,STIS observations 
show HH\,158 to be at a P.A. of 223\degr\ with respect to DG Tau \citep{Bacciotti02} and that the 
blueshifted HH\,158 ``micro\,-\,jet'' can be traced to within $\sim$ 0\farcs1 from the star 
\citep{Bacciotti00}. Larger scale studies show the HH\,158 outflow to extend to $\sim$ 11\arcsec, with 
a number of resolved knots \citep{Eisloffel98}. A redshifted jet $\sim$ 1\arcsec\ in length is detected 
$\sim$ 0\farcs45 from the source \citep{Lavalley97}. There is a  semi\,-\,circular nebula associated 
with DG Tau \citep{Mundt87} which HST data resolves as three nested arcs of reflection at 2\arcsec$\!$,
5\arcsec\ and 10\arcsec\ from DG Tau \citep{Stapelfeldt97}. The axis of symmetry of all three arcs is 
$\sim$ 225\degr\ with respect to DG Tau which is well aligned with the P.A. of HH\,158, suggesting that
these arcs are the illuminated edges of a cavity carved by the blueshifted jet.  

DG Tau B is located 47\arcsec\ south and 28\arcsec\ west of DG Tau and is a low luminosity {\em Class 
I} source with \Lstar = 0.88\Lsun \citep{Jones86}. It is the driving source of an optical jet (HH\,159) 
also found by \cite{Mundt83} and seen to be bipolar by \cite{Jones86}. HST images have resolved DG Tau 
B as a compact bipolar reflection nebula with no optically visible star \citep{Stapelfeldt97}. The 
eastern lobe of the reflection nebula is `V' shaped and the P.A. of its axis of symmetry is 122\degr\ 
\citep{Padgett99}. The 56\arcsec\ redshifted jet and 15\arcsec\ blueshifted jet have been observed both 
in imaging mode \citep{Mundt83, Mundt91,Eisloffel98} and spectroscopically 
\citep{Jones86,Mundt87,Eisloffel98}. Position angles of 122\degr\ for the blueshifted jet and 296\degr\ 
for the redshifted jet were determined by \cite{Mundt91}. However examination of our images would 
suggest that the P.A. of the blueshifted jet is closer to 116\degr\ i.e. the two jets are diametrically 
opposed as one might expect. In either case, the blueshifted jet is approximately coincident with the 
axis of symmetry of the reflection nebula, suggesting that the `V' delineates the walls of a cavity 
cleared by it. The redshifted jet passes through the centre of a well collimated redshifted CO emission 
lobe \citep{Mitchell94,Mitchell97}.

The most spectacular object in this outflow is HH\,702 - an $\sim$ 4\arcmin\ long shock system to 
the southwest of DG Tau. This object has already been noted by \cite{Sun03}, however our H$\alpha$     
images show more detail than the [SII] image of \cite{Sun03}. HH\,702 consists of five bright knots 
(Fig.\ \ref{dgtau_all}d) (one of which (E) was not noted by \cite{Sun03}), two fainter knots (F and G) 
and a trail of emission to the northeast. From our images, the trail of emission is first seen 7\farcm9
from DG Tau at a P.A. of 218\degr, pointing back towards DG Tau, and stretches a distance of 2\farcm75 
to Knot A. Knot A is an intense emission knot at 10\farcm6 from DG Tau, with a star directly to the 
east. Knots C and D are on either side of a star and can be seen more clearly in the continuum 
subtracted image inset in Fig.\ \ref{dgtau_all}d.  While a number of IRAS sources in the vicinity have 
been previously suggested as possible sources for HH\,702 \citep{Sun03}, we think it is more likely 
that this object is driven by DG Tau given the alignment between HH\,702 and HH\,158. The most distant 
optically visible object in the blueshifted outflow, Knot G of HH\,702 is 12\farcm32 from DG Tau, which
is 0.5pc in projected length at a distance of 140pc to the Taurus\,-\,Auriga cloud.

Our observations reveal a new HH complex (HH\,830) in the DG Tau outflow, details of which are given in 
Table\ \ref{dgtau_objects}. HH\,830 is a faint complex 9\farcm6 to the northeast of DG Tau and consists 
of three separate emission objects A -- C (Fig.\ \ref{dgtau_all}c). HH\,830\,A is a 44\arcsec\ long 
linear object orientated east\,/\,west and dips southwards towards its western edge. HH\,830\,B is 
$\sim$ 11\farcm3 from DG Tau and is brighter than HH\,830\,A but is much smaller in size 
(Fig.\ \ref{dgtau_all}b). HH\,830\,C is the most distant optically visible object in this outflow and 
appears to be two separate emission regions at 14\farcm1 (at P.A. of 47\degr) and 14\farcm4 (at P.A. of 
50\degr) from the source (Fig.\ \ref{dgtau_all}b). This object is at one edge of our image and 
there there might be more emission further east. The HH\,830 knots are comparable in H$\alpha$ and 
[SII] emission. Both HH\,830\,A and the centre of HH\,830\,C are along a P.A. of $\sim$ 48\degr\ with 
respect to DG Tau. Thus they are reasonably well aligned with the redshifted jet which we estimate to 
be at 44\degr\ from the images of \cite{Lavalley97}. HH\,830\,B is slightly off this axis, at a P.A. of
52\degr. The total length of the redshifted outflow is 14\farcm4 i.e. 0.59pc, giving the total 
projected length of the DG Tau bipolar outflow to be 1.09pc.

The outflow direction changes by 5\degr\ in the blueshifted outflow and 8\degr\ in the redshifted 
outflow giving the extended outflow a `C' shaped morphology, with DG Tau at the apex of the `C'. 
Similar morphologies have been noted in a small number of large\,-\,scale outflows from less evolved 
sources, for example the HH\,366 outflow in Barnard 5 \citep{Bally96}. This morphology may be analogous
to the head\,-\,tail extragalactic radio sources where a curved morphology is created between the radio 
source (head) and the bipolar jet (tails) due to the motion of the source through the ISM 
\citep{Edge95,Valentijn81}. In the YSO case, typical stellar velocities of around 5 -- 10 kms$^{-1}$ 
with respect to the parent cloud, combined with jet velocities of approximately 150kms$^{-1}$, might 
produce a similar effect. 

\begin{table}
\begin{tabular}{llccc}
\hline \hline
Object      &Source     &Angular              &P.A.$^b$            &Spatial        \\ 
            &           &Separation$^a$       &/\degr           &Extent$^c$     \\ \hline
HH\,830\,A  &DG Tau     &9\farcm6             &48              &5\arcsec $\times$ 44\arcsec     \\
HH\,830\,B  &DG Tau     &11\farcm3            &48              &2\arcsec $\times$ 2\arcsec  \\ 
HH\,830\,C  &DG Tau     &14\farcm1            &47              &21\arcsec $\times$ 9\arcsec \\
            &           &14\farcm4            &50              &6\arcsec $\times$ 3\arcsec \\
HH\,836\,A  &DG Tau B   &2\farcm6             &116             &14\arcsec $\times$ 7\arcsec \\
HH\,836\,B  &DG Tau B   &4\farcm4             &122             &12\arcsec $\times$ 3\arcsec \\
HH\,837     &DG Tau B   &10\farcm4            &117             &21\arcsec $\times$ 32\arcsec $^d$  \\
            &           &10\farcm4            &117             &19\arcsec $\times$ 18\arcsec $^e$  \\
HH\,838     &\ \ \ \ \ \ \ ?  &               &                &43\arcsec $\times$ 4\arcsec \\
HH\,839     &\ \ \ \ \  \ \ ?$^f$ &           &126             &3\arcsec $\times$ 10\arcsec \\ \hline
\end{tabular}
\caption{Angular separations, P.A.s and spatial extent of newly discovered HH objects in the DG Tau and 
DG Tau B region.
\newline $^a$ : Angular distance from the presumed source.
\newline $^b$ : P.A. with respect to the presumed source.
\newline $^c$ : Width and length of the object respectively.                          
\newline $^d$ : The morphology of HH\,837 is different in H$\alpha$ and [SII] emission; this is the 
width and length of this object in H$\alpha$.
\newline $^e$ : The morphology of HH\,837 is different in H$\alpha$ and [SII] emission; this is the 
width and length of this object in [SII]. 
\newline $^f$ : The source of this outflow appears to be a stellar object at its base (see 
Section~\ref{sec-dgtau}).}
\label{dgtau_objects}
\end{table}

\begin{figure}
\resizebox{\hsize}{!}{\includegraphics{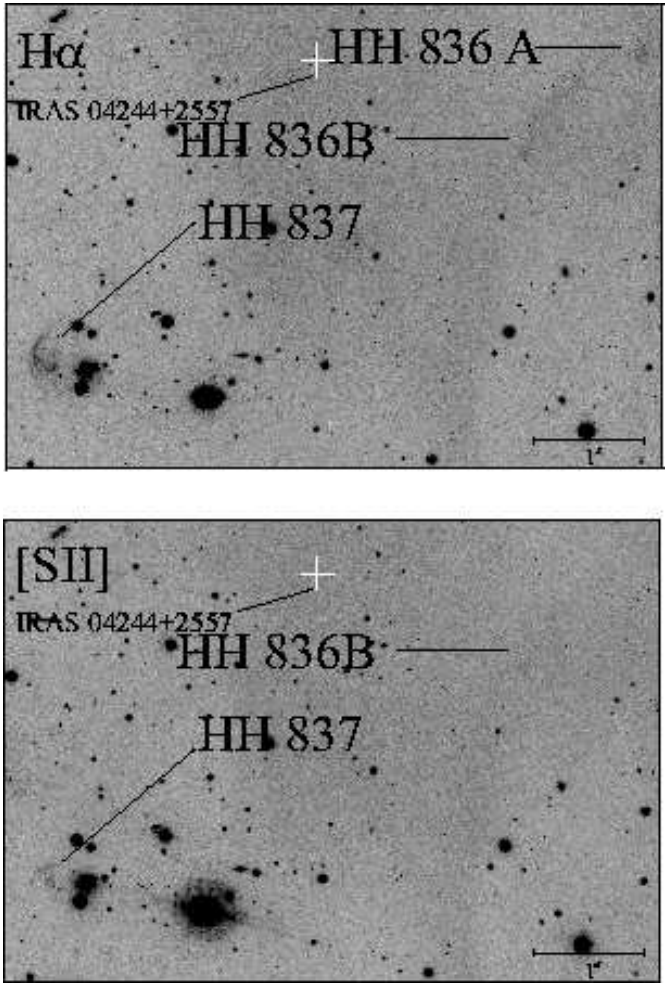}}
\caption{{\it HH\,836 and HH\,837 :\/} The objects to the southeast of DG Tau B are shown here in both
H$\alpha$ and [SII]. See Fig.\ \ref{dgtau_all}a for the extended DG Tau B outflow. HH\,836\,A is only 
seen in H$\alpha$, and HH\,836\,B is stronger in H$\alpha$ than in [SII]. There is a distinct 
difference between HH\,837 in H$\alpha$ and [SII]. The nearby IRAS source, which is also marked on 
Fig.\ \ref{dgtau_all}a, is marked here.}
\label{dgtauB_hh912_913}
\end{figure}

Our observations also reveal two faint new objects to the southeast of DG\,Tau\,B, with a third more
evident one further out in the same direction (Figs.\ \ref{dgtau_all} \& \ref{dgtauB_hh912_913}), 
details of which are given in Table\ \ref{dgtau_objects}. HH\,836\,A is at $\sim$ 2\farcm6 from the 
source and is seen only in H$\alpha$. HH\,836\,B is at 4\farcm4 from DG Tau B and is seen in both 
H$\alpha$ and [SII]. A trail of emission appears to curve back towards HH\,836\,A from HH\,836\,B in the
H$\alpha$ image. These HH objects are quite faint and diffuse. The most distant optically visible
object in the blueshifted outflow is HH\,837 at 10\farcm4 from DG Tau B. In [SII] this object appears 
to have a `V' shaped morphology pointing eastwards. In H$\alpha$ it is much stronger and is clearly bow 
shaped with some emission over\,-\,lapping the [SII] emission to the west of the bow 
(Fig.\ \ref{dgtauB_hh912_913}). There is no indication of this bow shock in the [SII] emission. Both 
HH\,836\,A and HH\,837 are at a P.A. of 116\degr\ and 117\degr\ respectively with respect to DG Tau B
and are well aligned with the blueshifted HH\,159 jet. HH\,836\,B is slightly off this axis at 122\degr\
suggesting that the direction of the blueshifted outflow has only varied slightly over the 10\farcm4 
distance from the source. Although IRAS 04244+2557 is a possible driving source for some of these HH 
objects (see Fig.\ \ref{dgtau_all}) it seems more than coincidental that all three are aligned with the
DG Tau B jet. The length of this outflow from the redshifted HH\,159 jet to HH\,837 is 11\farcm3 which 
is a projected length of 0.46pc. It is interesting to note that the opening angle for the blueshifted 
jet was initially estimated to be $\sim$ 17\degr\ i.e. almost four times that of the redshifted jet at 
$\sim$ 4.5\degr\ \citep{Mundt91}. Our images of the extended blueshifted outflow show that it has 
clearly recollimated; the extended opening angle (to HH\,837) is now 3\degr.

Two other HH objects are found in this region which do not appear to be related to either the DG Tau 
or the DG Tau B outflows. HH\,838 is 1\farcm6 at a P.A. of 286\degr\ from DG Tau B and is a faint, 
43\arcsec\ by $\sim$ 4\arcsec\ linear HH object running north\,/\,south (Fig.\ \ref{dgtau_all}) seen 
only in H$\alpha$ emission. All known IRAS sources in the regions are marked on Fig.\ \ref{dgtau_all} 
but none of these appear to be a likely driving source for HH\,838. Our images do not show any other HH
emission aligned with HH\,838. The second HH object, HH\,839 to the east of DG Tau B, consists of three
``knots''  (seen clearly in Fig.\ \ref{dgtau_all}e), of which the most northwestern one appears to be a
mixture of continuum and HH emission and is probably the location of the source driving the outflow. 
The other two knots which are at a P.A. of 126\degr\ with respect to this assumed source appear to be 
an $\sim$ 10\arcsec\ long outflow and are equally strong in H$\alpha$ and [SII]. This outflow is 
approximately  parallel to the DG Tau B outflow. HH\,839 is quite bright in our I band images so it 
possibly contains some reflection nebulosity as well. We can find no other obvious driving source for 
this outflow.

\subsection{The DO Tau and HV Tau region}
\label{sec-DOTAU}

\begin{figure*}
\resizebox{\hsize}{!}{\includegraphics{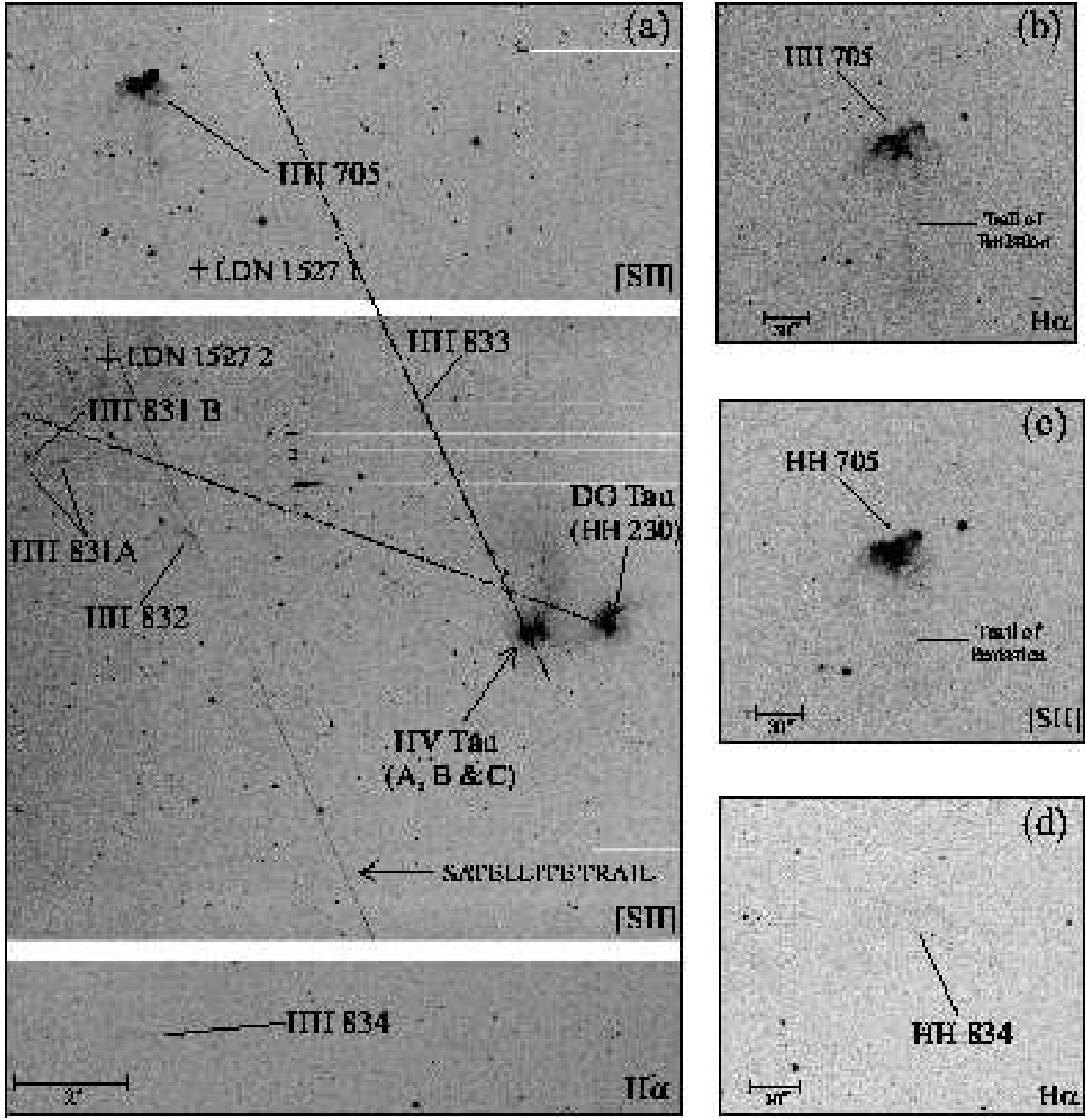}}
\caption{{\it DO Tau\,/\, HV Tau :\/} (a) Mosaic of the DO Tau and HV Tau C regions showing the newly 
discovered HH objects HH\,831 -- HH\,834. The positions of candidate sources for these HH objects are 
marked. HH\,230 is not seen in our optical images because of the DO Tau nebulosity, however the dashed 
line from DO Tau is at 70\degr\ marking the direction of the redshifted jet. There is also a dashed 
line at 25\degr\ from HV Tau C marking the P.A. of its blueshifted ``micro\,-\,jet''. HH\,831 and 
HH\,832 can be seen more clearly in Figs.\ \ref{dotau_hh906_907} and\ \ref{dotau_hh906}. (b) HH\,705 in
H$\alpha$ emission. (c) HH\,705 in [SII] emission. (d) HH\,834 in H$\alpha$. Note that HH\,834 is not 
seen in [SII] emission.}  
\label{dotau_flow}
\end{figure*}
 
DO Tau is a CTTS associated with an arc\,-\,like nebula. A bipolar jet (HH\,230) extending for 
approximately 2\arcsec\ to 4\arcsec\ was observed spectroscopically from DO Tau by \cite{Hirth94} 
with the redshifted jet at a P.A. of $\sim$ 70\degr\ \citep{Hirth94}. DO Tau is an M0 star 
\citep{Herbig88} with a mass in the range 0.3 -- 0.7 \Msun and an age of about 1.6 $\times$ 10$^5$ years
\citep{Beckwith90,Hartigan95}. To the east of DO Tau is the triple system HV Tau consisting of a close 
binary, A and B \citep{Simon96}, and a third component C $\sim$ 4\arcsec\ northeast of the close binary.
HV Tau C is an actively accreting CTTS \citep{Woitas98} and a bipolar ``micro\,-\,jet'' $\sim$ 1\farcs5 
in length has recently been observed emerging from it \citep{Stapelfeldt03}. We have estimated the P.A. 
of its blueshifted jet to be at 25\degr\ with respect to the source from the images of 
\cite{Stapelfeldt03}. 

Five HH objects were found in this region and are marked on Fig.\ \ref{dotau_flow}a with details given 
in Table\ \ref{dotau_objects}. Neither the DO Tau nor the HV Tau C jets are seen in our optical 
images because of the bright nebulosity around both stars. HH\,831 is found roughly northeast of DO Tau 
and is seen in both [SII] and H$\alpha$ although there are morphological differences between these two 
images. HH\,832 and HH\,833 are also seen to the northeast of DO Tau however they are only visible in 
[SII]. HH\,834 is the only object to the southeast and is seen only in H$\alpha$.

\begin{figure}
\resizebox{\hsize}{!}{\includegraphics{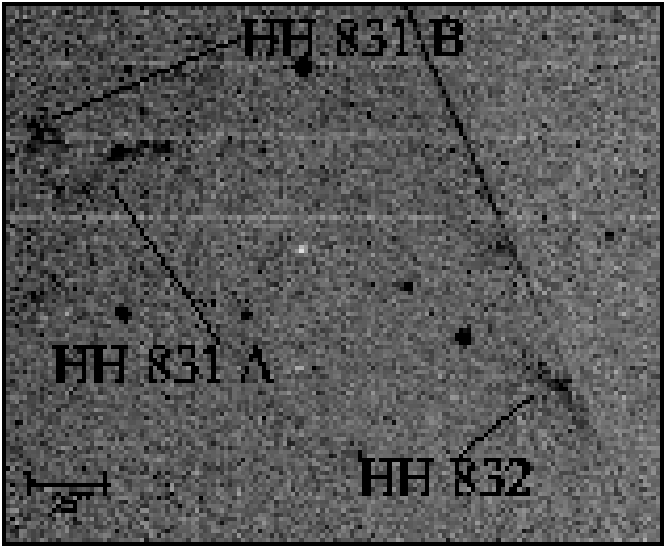}}
\caption{{\it HH\,831 and HH\,832 [SII] :\/} [SII] image of the newly discovered objects HH\,831 and 
HH\,832. Note that the morphology of HH\,831 is quite different here to the H$\alpha$ 
image (Fig.\ \ref{dotau_hh906}) and that HH\,832 is clearly seen in [SII] emission.}
\label{dotau_hh906_907}
\end{figure}

\begin{figure}
\resizebox{\hsize}{!}{\includegraphics{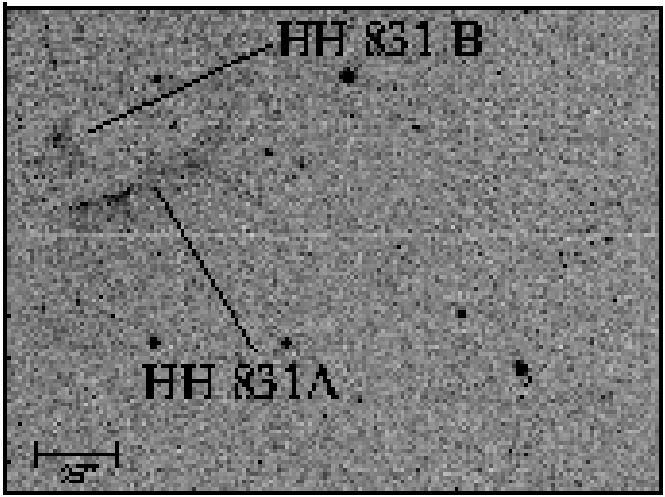}}
\caption{{\it HH\,831 H$\alpha$ :\/} HH\,831 as seen in H$\alpha$ emission. Note that 
this object is much fainter in H$\alpha$ emission than in [SII] (Fig.\ \ref{dotau_hh906_907}) and there
is no indication of HH\,832 here.}
\label{dotau_hh906}
\end{figure}

In H$\alpha$ images (Fig.\ \ref{dotau_hh906}) HH\,831\,A is a 52\arcsec\ long structure consisting of
three small joined arcs and is fainter towards the northwest and HH\,831\,B is a diffuse knot. In [SII] 
images (Fig.\ \ref{dotau_hh906_907}) HH\,831\,A is seen as two separate emission regions, the most 
distant of which is at 10\farcm8 from DO Tau. HH\,831\,B contains a number of bright emission regions 
in [SII], more than are seen in H$\alpha$ and is at a distance of 11$'$ from DO Tau. HH\,831\,B is only
$\sim$ 36\arcsec\ from the eastern edge of our field of view so it is very possible that there is more  
emission in this direction. HH\,831 is at a P.A. of $\sim$ 74\degr\ with respect to DO Tau and is well 
aligned with the redshifted jet. 

HH\,832 (Fig.\ \ref{dotau_hh906_907}) is seen only in [SII] and is at 78\degr\ and 7\farcm7 from DO 
Tau. It is an $\sim$ 6\arcsec\ long linear object with faint emission to the northeast. There is a 
nearby trail from a passing satellite.  

HH\,833 (Fig.\ \ref{dotau_flow}a) is a diffuse region 4\farcm6 from HV Tau C at a P.A. of 25\degr\ and 
is seen only in [SII]. HH\,705, to the north of HH\,831, is a complex object containing a number of 
bright emission regions (Figs.\ \ref{dotau_flow}b and c). This object was discovered by \cite{Sun03} 
however our images show it to be more intense and complex than their [SII] images. This object may be a 
bow shock, with faint emission trailing to the southeast and southwest forming the wings and there is a 
1\farcm3 long trail to the south. Its total width, as measured between the edges of the wings, is $\sim$
1$'$ and its total length is 1\farcm9. HH\,834 is very faint and seen only in H$\alpha$ 
(Fig.\ \ref{dotau_flow}d). It is linear and has a total length of 42\arcsec.

\begin{table}
\begin{tabular}{llccc}
\hline \hline
Object      &Source        &Angular              &P.A.$^b$  &Spatial        \\       
            &              &Separation$^a$       &/\degr     &Extent$^c$     \\ \hline
HH\,831\,A  &DO Tau        &10\farcm8            &74        &4\arcsec $\times$ 52\arcsec     \\
HH\,831\,B  &DO Tau        &11\arcmin\           &74        &14\arcsec $\times$ 8\arcsec     \\ 
HH\,832     &DO Tau        &7\farcm7             &78        &4\arcsec $\times$ 29\arcsec      \\
HH\,833     &HV Tau C      &4\farcm6             &25        &10\arcsec $\times$ 8\arcsec \\
HH\,834     &\ \ \ \ \ \ ? &                     &          &42\arcsec $\times$ 6\arcsec  \\ \hline
\end{tabular}
\caption{Angular separations, P.A.s and spatial extent of newly discovered HH objects in the DO Tau and 
HV Tau C region.
\newline $^a$ : Angular distance from the presumed source.
\newline $^b$ : P.A. with respect to the presumed source.
\newline $^c$ : Width and length of the object respectively.}                          
\label{dotau_objects}
\end{table}

The locations of these new objects in relation to previously known outflows suggests possible links. 
HH\,230 is at $\sim$ 70\degr\ with respect to DO Tau so it is likely that HH\,831 at 74\degr, and 
possibly HH\,832 at a P.A. of 78\degr, are part of the same outflow. There is a gap of $\sim$ 7\farcm7 
(0.43pc) between the 4\arcsec\ redshifted HH\,230 jet and HH\,832 where no HH emission is seen. If 
HH\,831 and HH\,832 are driven by DO Tau then the length of the redshifted outflow is $\sim$ 11\farcm07 
(0.45pc). 

Considering its alignment with the previously known ``micro\,-\,jet'' from HV Tau C, it is most likely 
that HH\,833 is driven by this source. HH\,705 could also be part of this outflow. Both the blueshifted 
jet from HV Tau C and HH\,833 are at 25\degr\ with respect to HV Tau C whilst HH\,705 is at $\sim$ 
36\degr\ with respect to this source at a distance of 12\arcmin\ from it. HH\,705 is $\sim$ 7\farcm7 
from HH\,833 and it is possible that the outflow direction could have changed by 11\degr\ over this 
distance. The southern trail of emission from HH\,705 might suggest that its driving source is situated 
to its south rather than southwest. In this case the radio sources LDN\,1527\,1 or LDN\,1527\,2 in 
LDN\,1527 \citep{Anglada92} are possible driving sources (as suggested by \cite{Sun03}, however it is 
unlikely that LDN\,1527\,1 could generate such a large object so nearby. We think HV Tau C is the most 
likely source of this object. Assuming HH\,833 and HH\,705 are driven by HV Tau C, the projected length
of this outflow is 0.49pc (12\arcmin). At present there is no obvious driving source for HH\,834. 

\subsection{RW Aur}
\label{sec-RWAUR}
RW Aur is a hierarchical triple system, with component A about 1\farcs4 at a P.A. of 76\degr\ from 
the close binary B and C \citep{Ghez93}. Both RW Aur A and B are actively accreting CTTSs 
\citep{Duchene99}. Although it was suggested as early as 1986 that RW Aur A itself might be a close 
binary \citep{Hartmann86} it is only in recent years that this has been confirmed with evidence of a  
low\,-\,mass secondary companion, possibly a brown dwarf \citep{Gahm99,Petrov01}. The HH\,229 bipolar 
jet is the only indication of an outflow in the vicinity of RW Aur to date \citep{Hamann94,Hirth94}. 
Furthermore, long slit spectroscopy confirms RW Aur A as the source of this outflow 
\citep{Hirth94,Hirth97} as does the 0\farcs1 resolution images of \cite{Dougados00}. Recently rotation 
has been observed at the base of the HH\,229 jet -- one of the first such outflows where rotation has 
been detected \citep{Coffey04}.
 
\begin{figure}
\resizebox{\hsize}{!}{\includegraphics{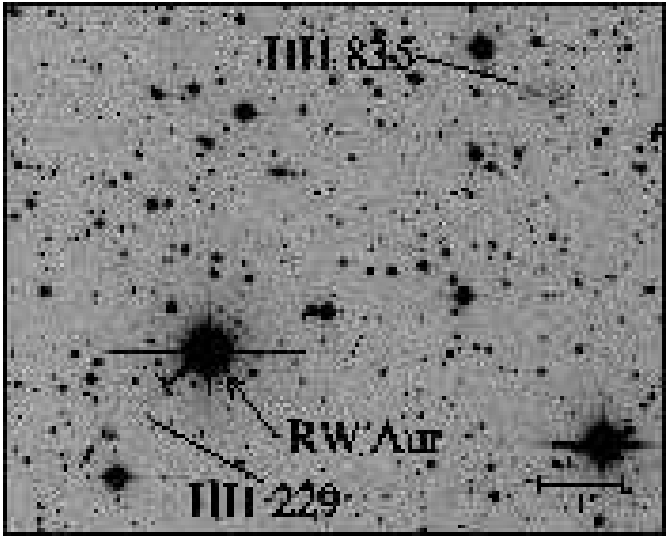}}
\caption{{\it RW Aur H$\alpha$ :\/} In this image of the RW Aur outflow, the HH\,229 
jet (and its direction) is clearly seen as is the newly discovered object HH\,835, which is seen in 
more detail in Fig.\ \ref{rwaur_hh911}.}
\label{rwaur_flow}
\end{figure}
                                             
\begin{figure}
\resizebox{\hsize}{!}{\includegraphics{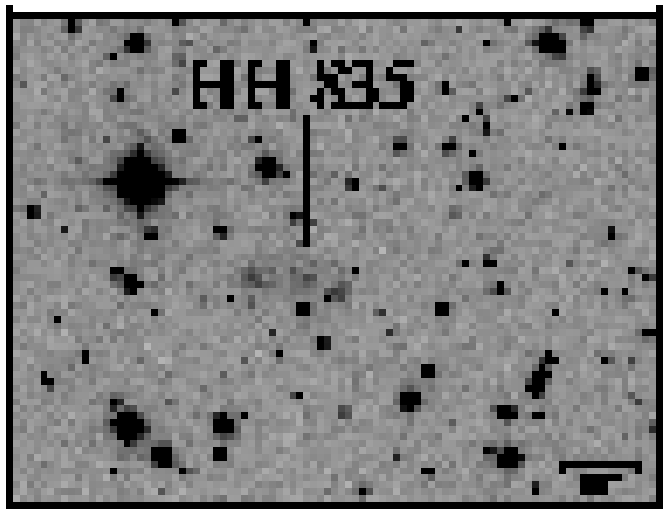}}
\caption{{\it HH\,835 H$\alpha$ :\/} HH\,835 to the northwest of RW Aur. We suggest that this is the 
east\,/\,west wing of a bow shock, with the north\,/\,south wing being much fainter in our 
optical images.}
\label{rwaur_hh911}
\end{figure}

The blueshifted component of the HH\,229 bipolar jet is at a P.A. of $\sim$ 130\degr\ with respect to 
RW Aur A \citep{Dougados00,Mundt98,Hirth97}. [SII] images of \cite{Mundt98} show the redshifted and 
blueshifted outflows to have lengths of 50\arcsec\ and 106\arcsec\ respectively, implying a total 
projected length of $\sim$ 0.1pc at a distance of 140pc. Both sides are well collimated, with a full 
opening angle for the redshifted jet of less than 6\degr\ \citep{Woitas02}. The detection of emission in
the redshifted jet can be traced back to 0\farcs1 from the source which implies an upper limit of 15 AU 
for the projected radius of its circumstellar disk.  

\begin{table}
\begin{tabular}{llccc}
\hline \hline
Object      &Source     &Angular               &P.A.$^b$       &Spatial        \\ 
            &           &Separation$^a$        &/\degr          &Extent$^c$     \\ \hline
HH\,835     &RW Aur     &5\farcm37             &310            &30\arcsec $\times$ 12\arcsec \\ \hline
\end{tabular}
\caption{Angular separations, P.A.s and spatial extent of the newly discovered HH object in the RW Aur  
outflow.
\newline $^a$ : Angular distance from the presumed source.
\newline $^b$ : P.A. with respect to the presumed source. 
\newline $^c$ : Width and length of the object respectively.}                          
\label{rwaur_objects}
\end{table}

Our images reveal a large object (HH\,835) 5\farcm37 to the northwest of RW Aur (See 
Fig.\ \ref{rwaur_flow} and Table\ \ref{rwaur_objects}). We suggest that this object is a bow shock with
only the $\sim$ 30\arcsec\ long northern wing visible in our H$\alpha$ images (Fig.\ \ref{rwaur_hh911}).
Given the position of HH\,835 on the known axis of the RW Aur outflow (P.A. of 310\degr\ with respect 
to RW Aur) we suggest that this object is part of the redshifted RW Aur A outflow. There is a distance 
of $\sim$ 4\farcm6 between the end of the redshifted jet and HH\,835 in which there is no optical 
evidence for outflow activity. The total observed projected length of the RW Aur outflow from the 
blueshifted jet to HH\,835 is 0.29pc (7$'$). 

\begin{table*}
\centering
\begin{tabular}{lcccllcl}
\hline \hline
Source          &L$_{bol}$    &Reference &Outflow Length &Associated Nebula    &\CS$^a$            &$\tau$$_{dyn}$  &$\theta$$_{flow}$$^b$   \\
                &/\,\Lsun     &     &/\,pc          &                     &\,/pc              &$\times$10$^4$ yr  &/\degr \\ \hline
DG Tau          &$\geq$7.6    &1    &1.09           &LDN\,1521            &7.4 $\times$ 5.4   &2.1             &10.6     \\  
DG Tau B        &0.88         &2    &0.46           &LDN\,1521            &7.4 $\times$ 5.4   &0.9             &1.8      \\  
CW Tau          &2.6          &3    &0.98           &LDN\,1495            &3.9 $\times$ 3.2   &2.0             &3.8      \\  
HV Tau C$^c$    &$\sim$0.5    &4    &0.49           &LDN\,1527$^d$        &0.03 $\times$ 0.01 &1.0             &4.1$^e$  \\
DO Tau$^f$      &3.5          &1    &0.45           &LDN\,1527$^d$        &0.03 $\times$ 0.01 &0.9             &3.1$^g$ \\  
RW Aur          &$\la$1.1$^h$ &5    &0.29           &Anonymous            &                   &0.6             &4        \\  \hline
\end{tabular}
\caption{Outflow lengths, dynamical timescales and degree of collimation for the five newly discovered 
parsec\,-\,scale outflows from (Class II) CTTSs and for the serendipitously discovered outflow from the 
Class I source DG Tau B. 
\newline$^a$ : Visual estimate of the approximate width and length of the associated nebula for each source.
\newline$^b$ : $\theta$$_{flow}$ is calculated by taking the width of the most distant shock in both the
red\,- and blueshifted outflows and dividing by the projected distances from the source. The mean value
of $\theta$$_{flow}$ for both outflows, where both are seen, is given here.
\newline$^c$ : Here we assume that HH\,833 and HH\,705 are part of the HV Tau C outflow (see
Section~\ref{sec-DOTAU}).
\newline$^d$ : The source is at the edge of this nebula.                                    
\newline$^e$ : This is the opening angle for the blueshifted outflow only -- the redshifted jet is too
short to measure accurately.
\newline$^f$ : Here we assume that HH\,230, HH\,831 and HH\,832 are part of the DO Tau outflow (see 
Section~\ref{sec-DOTAU}).
\newline$^g$ : This is the opening angle for the redshifted outflow only -- there is no optically 
visible blueshifted outflow (see Section~\ref{sec-DOTAU}).
\newline$^h$ : This is the luminosity of the RW Aur A binary and so is an upper limit for the binary 
component driving this outflow (see Section~\ref{sec-RWAUR}).
\newline References : 1. \cite{Cohen79}; 2. \cite{Jones86}; 3. \cite{GomezdeCastro93}; 4. \cite{Stapelfeldt03};
\newline 5. \cite{Petrov01}}
\label{parameters}
\end{table*}
\section{Discussion}

\label{sec-discussion}
We have shown that five of the seven CTTSs examined in the Taurus\,-\,Auriga Cloud (namely CW Tau, DG 
Tau, DO Tau, HV Tau C and RW Aur) drive outflows of the order of 1pc in length. We have estimated the 
length, dynamical timescale and degree of collimation for all six of the outflows discussed in 
Sections~\ref{sec-resultsI} and these are presented in Table\ \ref{parameters}. Here we examine the 
parameters and the morphological trends of these CTTS outflows in more detail, and compare them to 
large\,-\,scale outflows from more embedded Class I sources. These trends can be used to infer details 
about the source, the propagation of the outflow into the ambient medium, and the evolution of the 
outflows with time.

\subsection{Outflow Morphology}
\label{sec-morphology}
Typically many large\,-\,scale outflows from Class I sources are either `S' shaped with the driving 
source approximately in the middle of the `S' - for example the HH\,47 outflow \citep{Heathcote96}, 
HH\,34 outflow \citep{Bally94} and the PV Cephei outflow \citep{Reipurth97} to name but a few; or the 
rarer `C' shaped with the source at the apex of the `C' -  for example the HH\,366 outflow in 
Barnard 5 \citep{Bally96}. It can be seen from the images presented in Section~\ref{sec-resultsI} that 
many of the outflows from CTTSs show evidence for some variations in P.A. with time, the effect being 
most evident in the CW Tau outflow. This outflow delineates an inverted `S' shape centered on CW Tau 
with the northern part of the `S' more elongated than the southern part. The change in outflow 
direction is possibly due to precession of the outflow axis 
\citep{Raga01,Masciadri02}.
 
It has also been noted from large\,-\,scale outflows driven by less evolved sources that the size and 
complexity of the HH shocks increases with distance from the source. As the supersonic outflow 
propagates through the parent cloud it interacts with both the ambient medium and slower, previous 
ejecta producing the HH shocks observed. Many of these shocks fade quickly with time, it is only the 
strongest and largest shocks that remain. As a consequence of this, the gap between consecutive HH 
objects\,/\,complexes increases with distance as the majority of the more distant shocks quickly 
fade leaving just a few, larger shocks \citep{Reipurth01,Bally97}. As these extended shocks have 
undergone many interactions with their surroundings, and have had time to evolve, they will often be 
complex, chaotic objects rather than the simple knot\,-\,like structures seen closer to the young star. 
Most of the outflows from CTTSs presented here also demonstrate these trends. 

\subsection{Outflow Length}
\label{sec-lengths}
From Table\ \ref{parameters} we see that the length of the outflows from CW Tau, DG Tau, DO Tau and HV 
Tau C is of the order of $\sim$ 1pc. This represents an increase over the previously known values by a 
factor of 480 for the HV Tau C outflow, and factors of between 120 and 160 for the other three. The RW 
Aur outflow was previously known to be $\sim$ 2\farcm7 in length and the revised value is only a factor 
of $\sim$ 2.6 larger (7\arcmin). Outflows of the order of 1pc should readily be expected from Class II 
low mass stars, the typical age of which is 10$^6$ years. Even if an outflow has an average velocity of 
around 50 kms$^{-1}$ (i.e. the minimum velocity required to excite the shocks we see) then over 10$^6$ 
years it could extend up to 50pc.  

The field of view captured by the Wide Field Camera is $\sim$ 34\arcmin\ square which projects to a 
maximum detectable outflow length of 1.4pc at the distance of the Taurus\,-\,Auriga Cloud. Outflow 
morphology suggests that the distance between consecutive HH objects \ complexes increases exponentially
with distance from the source (Section~\ref{sec-morphology}) so if there is emission beyond our field of
view then it may easily be on scales of tens of parsecs, near the edges of the Taurus\,-\,Auriga Cloud 
boundaries, and well beyond the field of view of the WFC. 

The apparent dynamical timescales for these outflows is estimated in Table\ \ref{parameters}, assuming 
the most distant objects are moving at a velocity of 50 kms$^{-1}$. Values of around 10$^4$ years are 
derived thus we are only observing a fraction of the stars' outflow histories. 
 
Our observations suggest that, in all cases, the outflows have blown out of the associated clump. Many 
of the most distant objects in the outflows are at the clump edges while a few, most notably HH\,830 and
HH\,702 in DG Tau are clearly well beyond. There is little doubt based on statistically estimated 
lifetimes  that these outflows are even larger than our field of view, and thus our data supports the 
idea that they have blown out of their parent cloud. Large\,-\,scale outflows blowing out of their 
parent cloud is also a well documented occurrence in outflows from Class I low\,-\,mass sources, as our 
own observations of DG Tau B show. We have previously shown that this also occurs for outflows from 
intermediate\,-\,mass YSOs \citep{McGroarty04}. 

\subsection{Degree of Collimation}
\label{sec-collimation}
For the CTTSs outflows discussed here we find a degree of collimation of the order of $\sim$ 5\degr\ 
(see Col. 11 of Table\ \ref{parameters}). This value is in good agreement with the degree of 
collimation of large\,-\,scale outflows from more embedded Class I stars \citep{Mundt91}. Although our 
sample of outflows from Class II stars is small, this suggests that collimation remains high {\em even 
as the source evolves}.

\subsection{Have these CTTSs gone through the FU Orionis Phase?}
The morphology and location of the HH objects\,/\,complexes associated with the five CTTSs suggest long 
quiescent outflow phases punctuated by mass ejections of varying strength. According to this scenario, 
the most violent of the latter have given rise to the most distant (from the YSO), long\,-\,lasting and 
extended HH complexes that we see today. Reipurth \& Aspin (1997, and references therein) have 
suggested that these major ejection events occur when the parent star undergoes FU Orionis\,-\,type 
outbursts, or FUor events. Before addressing whether this is the case, and explaining why we think our 
sample, albeit small, lends considerable weight to the hypothesis, it is worthwhile saying a few words 
about the FU Orionis phenomenon. 

When a YSO undergoes an FUor outburst, its optical brightness increases by several orders of 
magnitude before decaying back to the pre-outburst luminosity over 50-100 years \citep{Hartmann96}. The
spectra of these so\,-\,called FUors in quiescence is that of a CTTS \citep{Hartmann96}, and there are 
at least nine known FUors \citep{Hartmann96} including the prototypical source -- FU Orionis itself 
\citep{Herbig77}. Almost immediately after their discovery it was suggested that FUor events 
could give rise to Herbig-Haro outflows \citep{Dopita78,Reipurth85}. More recently, as already stated, 
\cite{Reipurth97b} investigated this possible connection, basing their conclusions largely on a sample 
of embedded Class I sources. As it is accepted that in quiescence FUors are CTTSs it seems more 
appropriate to examine outflows of the latter for any ``fossil record'' of FU Orionis\,-\,type 
outbursts. In particular it is not obvious {\em apriori} that the outburst timescales for Class I and 
Class II stars are similar.

Examining Table\ \ref{parameters} we see that the dynamical timescales of the {\em extended} HH 
complexes, associated with our CTTSs, are typically 10$^4$ years. \cite{Herbig77} and \cite{Herbig03} 
have estimated the mean time between successive FUor outbursts is 10$^4$ years, suggesting a clear 
link. Moreover the relative number of FUors in comparison to CTTSs also supports this link.  

\section{Conclusions}
\label{sec-conclusions}
We have shown here that a number of (Class II) CTTSs, DG Tau, CW Tau, DO Tau, HV Tau C and RW Aur, which
were previously known to drive ``micro\,-\,jets'' or short outflows of  $\la$ 1$'$ ($\la$ 0.04pc at the 
distance of the Taurus\,-\,Auriga cloud) in length, are actually capable of driving outflows with 
lengths of the order of 1pc. The serendipitous discovery of a 0.5pc long outflow from the Class I source
DG Tau B is also reported here.

The morphological trends observed in the CTTSs outflows are comparable to those noted in younger sources
i.e. increasing distance between successive HH objects coupled with increased size and complexity with 
distance from the source. In a few cases, small variations in the direction of propagation of the 
outflow have been found. The high degree of collimation of the five extended outflows from CTTSs 
compared well with that observed in the case of extended large\,-\,scale outflows from less evolved 
sources, suggesting that outflows remain focussed even as the source evolves from the Class I to the 
Class II stage. It is clear that the observed parsec\,-\,scale lengths of the CTTS outflows are minimum 
values and in reality they are much larger. These outflows all show evidence for having blown out of the
parent cloud.  

The apparent dynamical timescale of these extended outflows is typically a few times 10$^4$ years. This 
suggests a linkage between the major accretion events that give rise to the largest HH complexes and 
the FU Orionis phenomenon. As FU Orionis stars are CTTSs in quiescence, the extended outflows of the 
latter provide the best ``fossil record'' to test this linkage. 

\acknowledgement{
We thank the anonymous referee for helpful comments that clarified the presentation of these results. 
FMcG and TPR acknowledge support from Enterprise Ireland. This research has made use of the SIMBAD 
database, operated at CDS, Strasbourg, France.}

\end{document}